\documentclass[12pt]{emulateapj}

\shorttitle{Epoch of disk settling}
\shortauthors{Kassin et al.}

\begin{document}

\title{The Epoch of Disk Settling: $z\sim1$ to Now}

\author{Susan A. Kassin,\altaffilmark{1,}\altaffilmark{2}
Benjamin J. Weiner,\altaffilmark{3} 
S. M. Faber,\altaffilmark{4}
Jonathan P. Gardner,\altaffilmark{1}
C. N. A. Willmer,\altaffilmark{3} 
Alison L. Coil,\altaffilmark{5} 
Michael C. Cooper,\altaffilmark{6,}\altaffilmark{7}  
Julien Devriendt,\altaffilmark{8}
Aaron A. Dutton,\altaffilmark{9}
Puragra Guhathakurta,\altaffilmark{4} 
David C. Koo,\altaffilmark{4}
A. J. Metevier,\altaffilmark{10}
Kai G. Noeske,\altaffilmark{11} and
Joel R. Primack\altaffilmark{12}}
\altaffiltext{1}{Astrophysics Science Division, Goddard Space Flight Center, Code 665, Greenbelt, MD 20771, susan.kassin@nasa.gov}
\altaffiltext{2}{NASA Postdoctoral Program Fellow}
\altaffiltext{3}{Steward Observatory, University of Arizona, Tucson, AZ}
\altaffiltext{4}{UCO/Lick Obs., Dept of Astronomy and Astrophysics,  University of California, Santa Cruz, CA 95064}
\altaffiltext{5}{Department of Physics, University of California, Santa Barbara, CA 93106}
\altaffiltext{6}{Center for Galaxy Evolution, Department of Physics and Astronomy, University of California, Irvine, 4129 Frederick Reines Hall, Irvine, CA 92697}
\altaffiltext{7}{Hubble Fellow}
\altaffiltext{8}{Department of Physics, Denys Wilkinson Building, Keble Road, Oxford, OX1 3RH, United Kingdom}
\altaffiltext{9}{Max Planck Institute for Astronomy, K\"onigstuhl 17, 69117, Heidelberg, Germany}
\altaffiltext{10}{Department of Physics and Astronomy, Sonoma State University, 1801 E. Cotati Ave., Rohnert Park, CA 94928}
\altaffiltext{11}{Space Telescope Science Institute, 3700 San Martin Drive, Baltimore, MD 21218}
\altaffiltext{12}{Department of Physics, University of California, 1156 High Street, Santa Cruz, CA 95064}

\begin{abstract}
We present evidence from a sample of 544 galaxies from the DEEP2 Survey for evolution of the internal kinematics of blue galaxies 
with stellar masses ranging $8.0 < {\rm log}\,M_* (M_{\odot}) < 10.7$ over $0.2<z<1.2$.  DEEP2 provides galaxy spectra and $Hubble$ imaging from which we measure emission-line kinematics and galaxy inclinations, respectively.  Our large sample allows us to overcome scatter intrinsic to galaxy properties in order to examine trends in kinematics.  We find that at a fixed stellar mass galaxies systematically decrease in disordered motions and increase in rotation velocity and potential well depth with time.  Massive galaxies are the most well-ordered at all times examined, with higher rotation velocities and less disordered motions than less massive galaxies.  We quantify disordered motions with an integrated gas velocity dispersion corrected for beam smearing ($\sigma_g$).  It is unlike the typical pressure-supported velocity dispersion measured for early type galaxies and galaxy bulges.  Because both seeing and the width of our spectral slits comprise a significant fraction of the galaxy sizes, $\sigma_g$ integrates over velocity gradients on large scales which can correspond to non-ordered gas kinematics.   We compile measurements of galaxy kinematics from the literature over $1.2<z<3.8$ and do not find any trends with redshift, likely for the most part because these datasets are biased toward the most highly star-forming systems.  In summary, over the last $\sim8$ billion years since $z=1.2$, blue galaxies evolve from disordered to ordered systems as they settle to become the rotation-dominated disk galaxies observed in the Universe today, with the most massive galaxies being the most evolved at any time.

\end{abstract}


\section{Introduction}
In the standard picture of disk galaxy formation, baryons enter into dark matter halos and dissipate to form disks at their centers 
\citep{rees,whit,fall,blum}.  The accretion of baryons onto galaxies is expected to be an on-going process, and it 
is not yet known how galaxy disks respond to it, or how they evolve as mass accretion rates change.  Only recently
have hydrodynamic simulations of galaxy formation been run with enough resolution to study this in detail 
for up to a few tens of galaxies \citep[e.g.][and references therein]{gove07, gove09, bour09,ceve,brook,kimm, mart}.  
Such simulations have yet to address the evolution of internal galaxy kinematics over a significant period of time, which is
the cleanest observational probe of disk galaxy assembly.  Internal kinematics tell us directly about the dynamical 
state of galaxies, reveal the potential well depths of individual galaxy-dark matter halo systems, and can be measured for 
a large sample of galaxies over a significant range in redshift.  

We have acquired a wealth of information about blue galaxies over the last $\sim8$ billion years since $z=1.2$ 
which can be roughly divided into two categories: those which imply a significant amount of evolution
and those which do not.  Among the observations which do not imply much evolution are those of galaxy 
luminosity functions.  Luminosity function studies find that the number density of blue galaxies is unchanging, 
and that blue galaxies fade by only $\sim 1$ $B$-band magnitude since $z\sim1$ \citep[e.g.,][]{will, fabe, bell}.
Similarly, studies of stellar mass functions of blue galaxies find no evolution to within 
uncertainties (\citealt{bund}; \citealt{borc}; \citealt{pozz}; although see \citealt{dror}).
In addition, galaxy sizes are only marginally smaller at $z=1$ compared to local galaxies \citep[by a factor of 1.4;][]{dutt}.
However, despite all this evidence for a slowly evolving population, there are some strong indications of substantial evolution.
Compared with blue galaxies of similar stellar masses today, those at $z\sim1$ have significantly 
higher star-formation rates by a factor of $\sim10$ \citep[e.g.,][]{kai1},
higher molecular gas fractions (for at least a few of the brighter galaxies which have been observed so far) by a factor of $\sim 2-4$ \citep{daddi,tacc},
and more disturbed morphologies \citep[e.g.,][]{abra96,abra01}.  In summary, blue galaxies do not
evolve much since $z\sim1$ in luminosity, stellar mass, or size, but they do evolve strongly in star-formation rate,
(likely) molecular gas fraction, and morphology.    In this paper, we attempt to understand this discrepancy and 
the mechanisms behind it by studying the internal kinematics of blue galaxies since $z\sim1$.


Due to the large intrinsic scatter in the kinematic properties of blue galaxies \citep{wei1, wei2, kass}, a sizable sample ($\ga100$) 
which is representative in terms of galaxy properties is  required to study internal galaxy kinematics over a significant look-back time.  
The first large study to address the kinematic evolution of blue galaxies since $z\sim1$ was \citet{wei1,wei2}
which used data for 1089 galaxies from the TKRS Survey.  Among other things, they found that about one-third to one-half of 
emission-line galaxies over $0.1 < z < 1.5$ have a significant or dominant component of disordered motions,
as measured via an integrated gas velocity dispersion ($\sigma_g$; as discussed in \S4.1).   \citet{wei1} proposed a new velocity indicator
to trace galaxy potential well depths.  It incorporates both rotation velocity ($V_{rot}$) and $\sigma_g$:
{\small $S_{K} \equiv \sqrt{KV_{rot}^2 + \sigma_g^2}$}.

\citet{kass} (hereafter K07) followed up on \citet{wei1,wei2}.  Their dataset consisted of 544 galaxies from the DEEP2 Survey
over $0.1<z<1.2$ with medium-resolution spectroscopy from which kinematics were measured and $Hubble/ACS$ imaging.
The $Hubble$ images were used to measure galaxy morphologies and axial ratios (which were used by K07
to correct rotation velocities for inclinations of galaxies to the line-of-sight).  The main results of K07 are:
\begin{itemize}

\item K07 showed that scatter in the stellar mass Tully-Fisher relation (TFR), i.e. the relation between galaxy
stellar mass and $V_{rot}$, is mostly to low $V_{rot}$ and is a strong function of morphology.  Galaxies which scatter to low $V_{rot}$ 
generally have disturbed or compact $Hubble$ morphologies.  Similarly, from samples of 33 and 68 galaxies in the 
IMAGES Survey at $z\sim0.6$, \citet{flor} and \citet{yang_images1}, respectively, found
a large scatter in the TFR due to galaxies with ``perturbed or complex kinematics."
Most previous studies of Tully-Fisher at both low and high redshift tried to
minimize scatter by selecting galaxies which are morphologically well-ordered disks (e.g., \citealt{verh}, \citealt{kass06}, \citealt{piza} and references
therein for low redshift, and e.g., \citealt{cons}, \citealt{mill} and references therein for $z\la1.5$).

\item K07 demonstrated that galaxies which scatter to low $V_{rot}$ in the TFR have a significant contribution to their
kinematics from disordered motions (as quantified by $\sigma_g$).  

\item K07 showed that when $V_{rot}$ and $\sigma_g$ are combined into a velocity indicator created to trace galaxy potential well depths
($S_{0.5}$ as defined above), a tight relation with stellar mass
results.   This relation is independent of galaxy morphology, non-evolving since $z=1.2$, and coincident with the Faber-Jackson
relation for early-type galaxies.  It demonstrates that galaxies are likely approximately virialized since $z\sim1$ in that
they have all the energy they will have today already at $z\sim1$.
This tightening of Tully-Fisher once $S_{0.5}$ is adopted was later found by 
other observational studies \citep{cres, puec_btf, lem15, lem3, verg, catin} and numerical simulations of interacting galaxies \citep{covi}. 

\item K07 found that the majority of scatter in the TFR at low redshift is at lower stellar masses ($<10^{10}$ M$_{\odot}$).
They found that the scatter at higher masses was low at low redshift, and increased with increasing redshift to $z=1.2$.  
K07 hypothesized that this meant higher mass galaxies have been settling onto the TFR since $z=1.2$, 
and that lower mass galaxies are settling onto the TFR today.  Similarly, studies with the IMAGES Survey argue for significant kinematic 
evolution since $z\sim0.6$ \citep[e.g.,][]{yang_images1, neic, puec_images3}.

\end{itemize}
The K07 study leaves a number of open issues:
\begin{itemize}

\item First, although it is clear that galaxies are evolving in 
$V_{rot}$ and $\sigma_g$ since $z=1.2$, this evolution is not quantified.

\item Secondly, it is also clear from K07 that kinematic evolution is a function of galaxy stellar mass, but this 
is not quantified or addressed in detail.

\item  Finally, although K07 found that the relation between $S_{0.5}$ and stellar mass relation does not evolve to within
uncertainties since $z=1.2$, it has yet to be determined whether galaxies evolve in $S_{0.5}$.

\end{itemize}

In this paper we address these open issues with a more sophisticated analysis of the dataset in K07.
In \S2 we detail how the K07 sample was selected; this was not included in K07 because it was a letter paper
and did not have the space to do so.  In \S 3 we discuss the stellar mass and $Hubble$ size measurements 
which are adopted from other studies, and describe the measurement of emission line extents
made in this paper.  In \S 4,  since it was not addressed in detail in K07, we describe how $V_{rot}$ and 
$\sigma_g$ are measured from emission lines in galaxy spectra and discuss what they measure.
We then show how $V_{rot}, \sigma_g,$ and $S_{0.5}$ evolve with redshift in \S 5, and discuss the influence of 
galaxy stellar mass and size on this evolution in \S 6.  In \S7 we quantitatively define a 
settled disk galaxy and study how the fraction of settled disk galaxies evolves with redshift.
Our findings are compared with previous measurements of galaxy kinematics in the literature over $0<z<3.8$ in \S 8.  
Conclusions are given in \S 9.
A $\Lambda$CDM cosmology is adopted throughout ($h=0.7, \Omega_m=0.3, \Omega_{\Lambda}=0.7$),
and all logarithms are base 10.

\begin{figure}    
\includegraphics[scale=0.85]{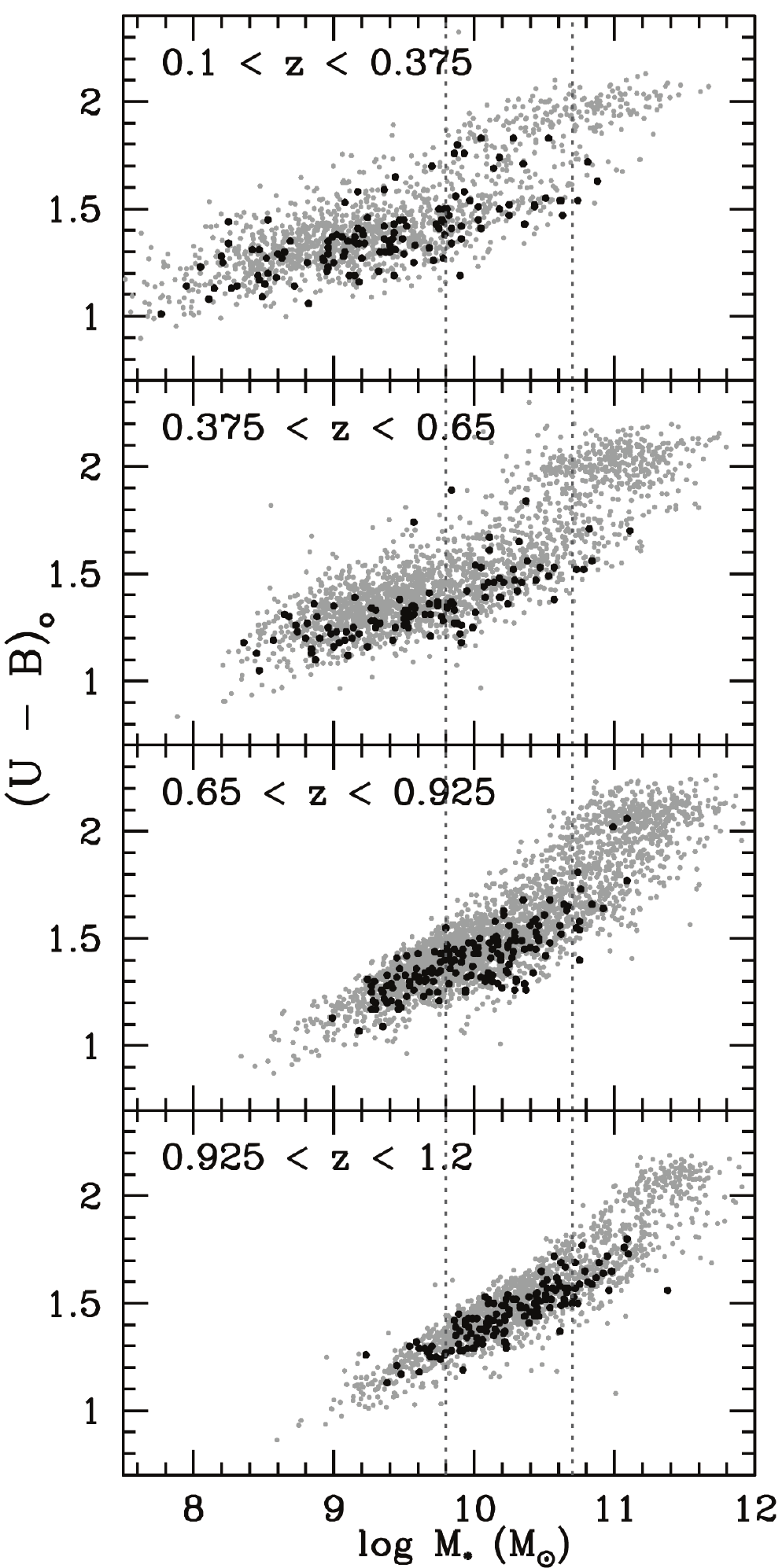}
\caption{Rest-frame $U-B$ color (AB system) versus stellar mass diagrams for galaxies with solid spectroscopic redshifts in field 1 of the DEEP2 Survey (grey points)
and for the K07 sub-sample which is used in this paper (black points) are shown in bins of redshift.  
The K07 sample provides good coverage over galaxy mass.
Although the K07 sample is moderately biased towards bluer galaxies at a fixed mass, the effect is not severe. 
Dotted lines demarcate the mass-limited sample used later in this paper.
\label{colormass}}
\end{figure}

\begin{figure*}
\includegraphics[scale=1.3]{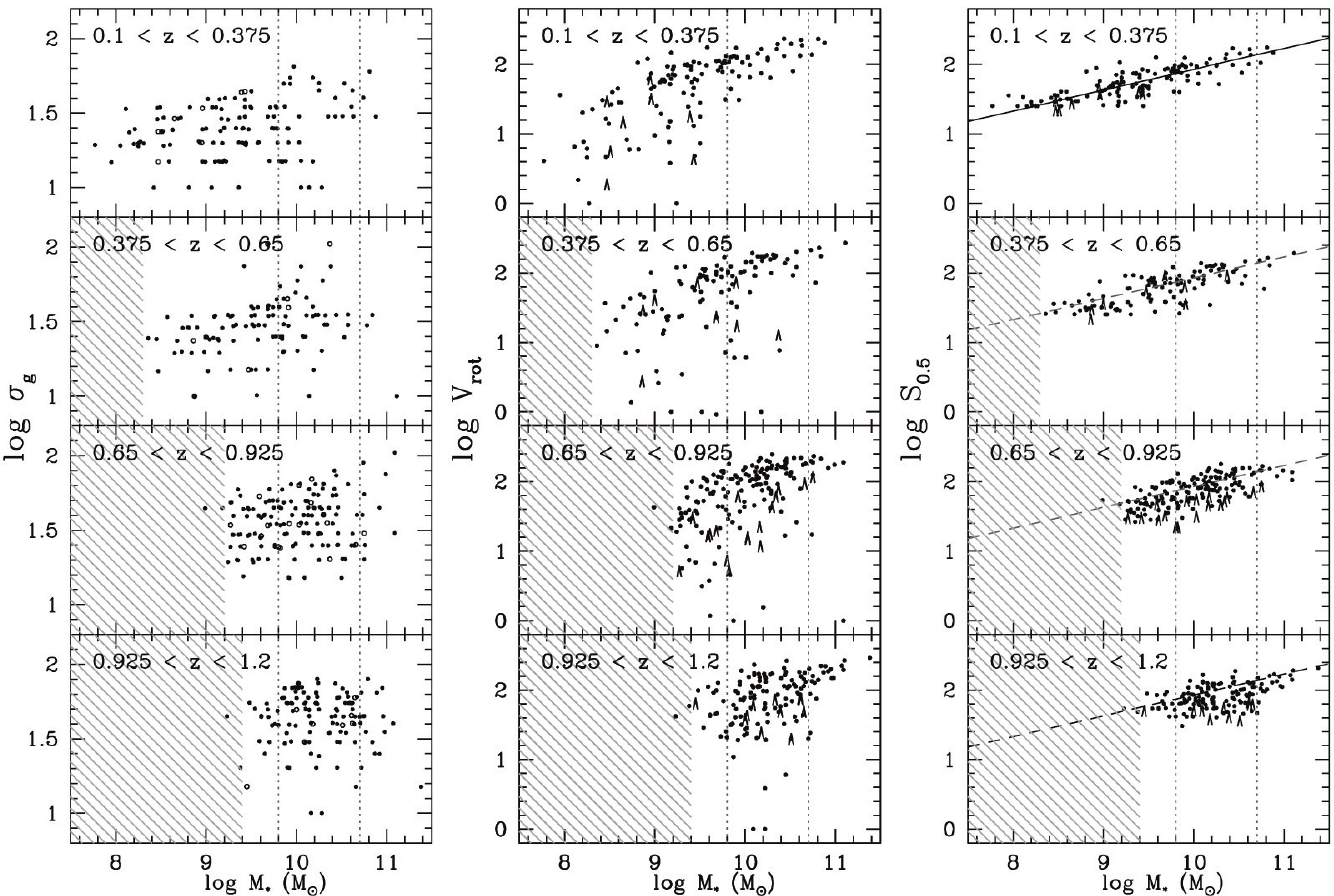}
\caption{Relationships between kinematic quantities ($\sigma_g$, $V_{rot}$, $S_{0.5}$) and stellar mass are shown in bins of redshift for the 
K07 sample.  Stellar masses are from \citealt{lihwai}.  The $V_{rot}$ and $S_{0.5}$ relations were shown previously in K07, and
we show the $\sigma_g$ relation here for completeness.  While large scatter is found for the $\sigma_g$ and $V_{rot}$ relations, 
the $S_{0.5}$ relation is significantly tighter and does not evolve to within uncertainties since $z=1.2$.  The fit to the $S_{0.5}$ 
relation for the lowest redshift bin is shown as a solid line (slope of $0.30\pm0.2$, intercept of $1.93\pm0.01$ at log $M_*$=10 M$_{\odot}$; K07),
and is repeated in the higher redshift bins as a dashed line.   It is consistent with the Faber-Jackson Relation for early type galaxies,
where $\sigma$ is measured from stellar absorption lines (K07).  Hashed regions denote areas where the survey
is not sensitive, and dotted lines demarcate the mass-limited sample used later in this paper.
If a galaxy is deemed too morphologically disturbed to accurately determine an inclination 
(9\% of the sample), $V_{rot}$ is not inclination corrected and the galaxy is shown as an upper limit symbol in the $V_{rot}$ and $S_{0.5}$ plots 
and an open circle in the $\sigma_g$ plot.
\label{tfr}}
\end{figure*}

\begin{figure*}
\includegraphics[scale=1.8]{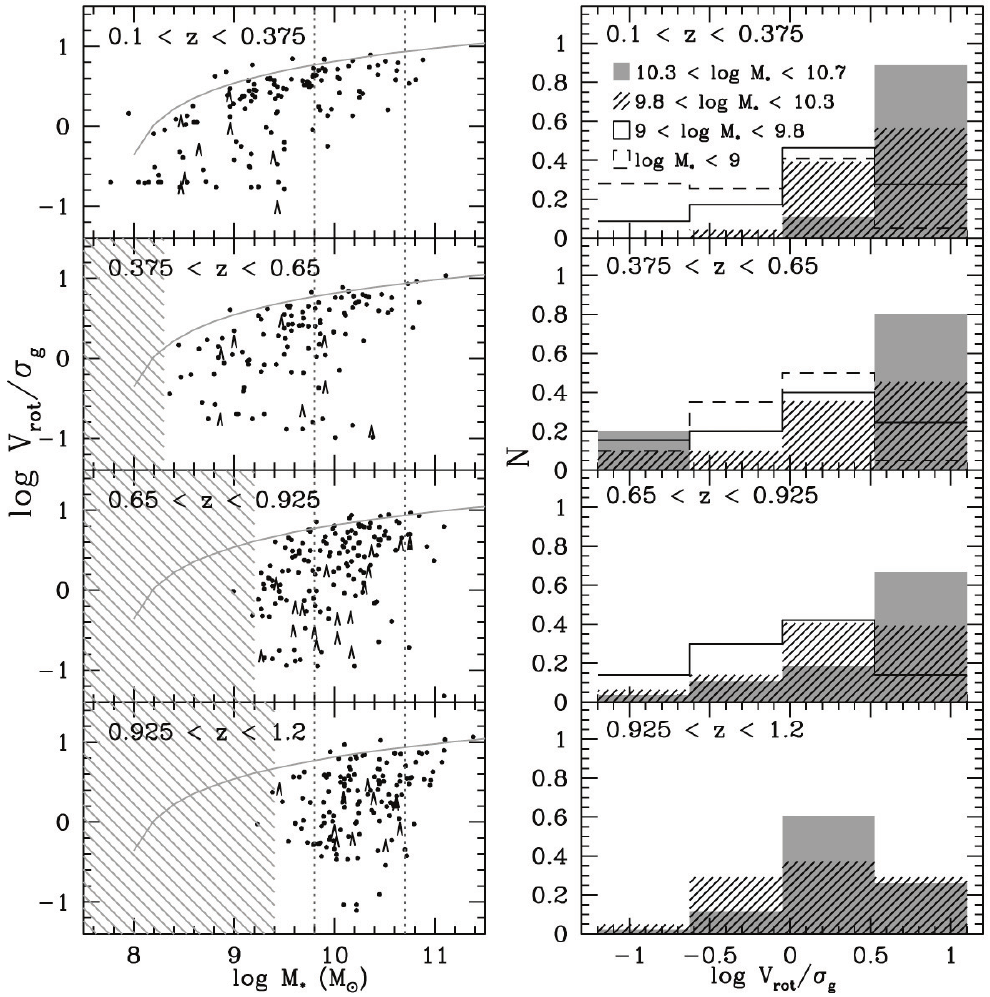}
\caption{Left: The ratio of ordered to disordered motions in a galaxy, $V_{rot}/\sigma_g$,
is shown as a function of stellar mass and in bins of redshift for the K07 sample.  Galaxy symbols are as in Figure~\ref{tfr}.  
Hashed regions denote areas where the survey is not sensitive, and dotted lines demarcate the mass-limited sample used later in this paper.
The galaxies define an upper envelope 
which does not evolve with redshift.  However, galaxies evolve within the envelope:  the lower-right corner evacuates
with time (from high to low redshift), with higher mass galaxies evolving the fastest.
Right:  This evolution is more clearly shown as histograms in $V_{rot}/\sigma_g$ for galaxies divided into stellar mass bins.
\label{velsig_mst}}
\end{figure*}

\newpage

\section{Sample Selection}

In this paper we use the dataset of K07.  Since K07 was a letter paper, it did not include a detailed description of 
how its galaxy sample was selected, so we include it here.  One of the main differences between K07 and most other 
studies of galaxy kinematics lies in its sample selection.
Typically, constraints on galaxy morphology are applied at the outset to remove disturbed and major-merger galaxies.
K07 instead included all galaxies with emission lines which are bright enough to measure 
kinematics from, regardless of morphology.  Since disturbed morphology and disordered motions (as measured by $\sigma_g$) are correlated, 
removing galaxies with disturbed morphologies would introduce a strong bias toward only the most 
rotation-dominated systems (K07).  Little or no evolution 
in internal galaxy kinematics would be found for a sample selected in this manner.  Moreover, such a sample would not be representative of 
galaxies at higher redshifts since they tend to have disturbed morphologies \citep[e.g.,][]{grif, wind, abra96,abra01}.   
Even when a strict morphological selection is loosened somewhat to include galaxies which are 
moderately disturbed but disk-like, still little scatter and no evolution in Tully-Fisher since $z=1.7$ are found \citep{mill,mill2}.
Furthermore, many studies of galaxy kinematics at medium to high redshift are limited to galaxies with extremely bright emission lines,
which are not representative of the general population of galaxies.
The sensitivity of DEEP2 allows the K07 sample to probe a large range in galaxy stellar mass and color (Figure~\ref{colormass}).

In addition to broad morphological selection criteria and a large range in stellar mass, 
in order to study the evolution of disk galaxy kinematics over a significant look-back time, 
medium resolution spectra of a {\it sizable} sample of galaxies ($\ga100$) which span a wide range in redshift are required.
A sizable sample is needed because emission-line galaxies have significant intrinsic scatter in their kinematic properties (\citealt{wei1, wei2}; K07).

The K07 sample derives from a very large collection of 9715 galaxies with solid spectroscopic redshifts from 
Keck/DEIMOS \citep{deimos} in field 1 of the DEEP2 Survey (May 2007 catalog), as described below. 
Field 1 is home to the AEGIS multi-wavelength survey \citep{aegis}. 
The limiting magnitude of DEEP2 is $R_{AB}<24.1$ and galaxies in field 1 are not subject to a color selection (\citealt{will}; Newman et al. submitted).  
In Figure~\ref{colormass} we show rest-frame color-magnitude diagrams for galaxies
in field 1 of DEEP2 and highlight the K07 sample.  The K07 sample provides good coverage over galaxy mass.
Although it is moderately biased toward blue galaxies at a fixed mass, the effect is not severe.

In the following we detail how galaxies were selected from field 1 of the DEEP2 Survey to be in the K07 sample.
First, although DEEP2 reaches a redshift of $z=1.5$, K07
chose to be conservative and limit their sample to $z=1.2$, leaving 8787 galaxies.
This helps to avoid the high-redshift end of the survey where the selection is in the rest-ultraviolet and
becomes dependent on color \citep{will}.  It also ensures that galaxy axis ratios, which are used to correct
rotation velocities for inclinations of galaxies to the line-of-sight, are measured from $Hubble$ 
imaging which has not yet shifted into the ultraviolet since these wavelengths
may not trace the bulk of the stellar mass of galaxies.  Next, galaxies were chosen to have 
{\it Hubble Space Telescope}/ACS imaging at $V$ and $I$, which further reduces the 
sample to 3523 galaxies.  This is because there are portions of field 1
which are not imaged by $Hubble$.  Aside from removing galaxies which did not have $Hubble$ imaging,
this requirement did not further affect our sample selection because the $Hubble$ images are deeper than the spectral survey.
For our most important selection, galaxies with bright enough emission lines to measure kinematics ($\ga 10^{-17}$ erg s$^{-1}$ cm$^{2}$) were chosen, leaving 1692 galaxies.

Galaxies were further selected by K07 to have inclinations measured from $Hubble$ images between
$30\degr < i < 70\degr$, and spectrographic slits with position angles aligned to within $40\degr$ of their major axes.
These cuts leave a sample of 755 galaxies.  Inclinations are measured from $V+I$-band $Hubble$ images 
using the {\tt SExtractor} software package \citep{sext}, as described in \citet{lotz}, and have an uncertainty of $\sim 10\degr$.
The inclination requirement avoids nearly face-on galaxies for which inclination measurements are very uncertain,
and minimizes the effects of dust in the determination of stellar mass for highly inclined systems. 
However, if the morphology of a galaxy was deemed by eye to be disturbed enough such that its inclination 
could not be reliably determined, it was included in the sample but not corrected for inclination.  There are only
24 of these galaxies and they are flagged in most of the following
analysis.  Removing them from the sample does not significantly affect the results of this paper.
For the slit position angles, due to the large slit width ($1$\arcsec) compared to the apparent sizes of galaxies in our sample ($\sim3$\arcsec)
and the effects of seeing, accurate kinematics can be measured if
slit position angles are offset by up to 40\degr\ from galaxy major axes \citep[Figure 13 of][]{wei1}. 
Finally, if an emission line was affected by a sky line or an instrumental artifact,
the corresponding galaxy was removed from the sample, leaving a total of 544 galaxies in the K07 sample.
No constraints on morphology were applied. 

To study the evolution of galaxy kinematics for a sample of blue galaxies which is mass-limited at all
redshifts, in \S5.1 we limit the K07 sample to galaxies with stellar masses over $9.8 < {\rm log}\,M_* (M_{\odot}) < 10.7$,
as demarcated in Figures~\ref{colormass}--~\ref{velsig_mst}.
Since the stellar mass function of blue galaxies does not evolve significantly since 
$z\sim1$ \citep{borc, bund, pozz}, we do not change this stellar mass range with redshift. 
This cut removes 274 galaxies from the K07 sample, leaving 270. 
Throughout the paper we will refer to this as the ``mass-limited sample."

\begin{figure}
\includegraphics[scale=0.85]{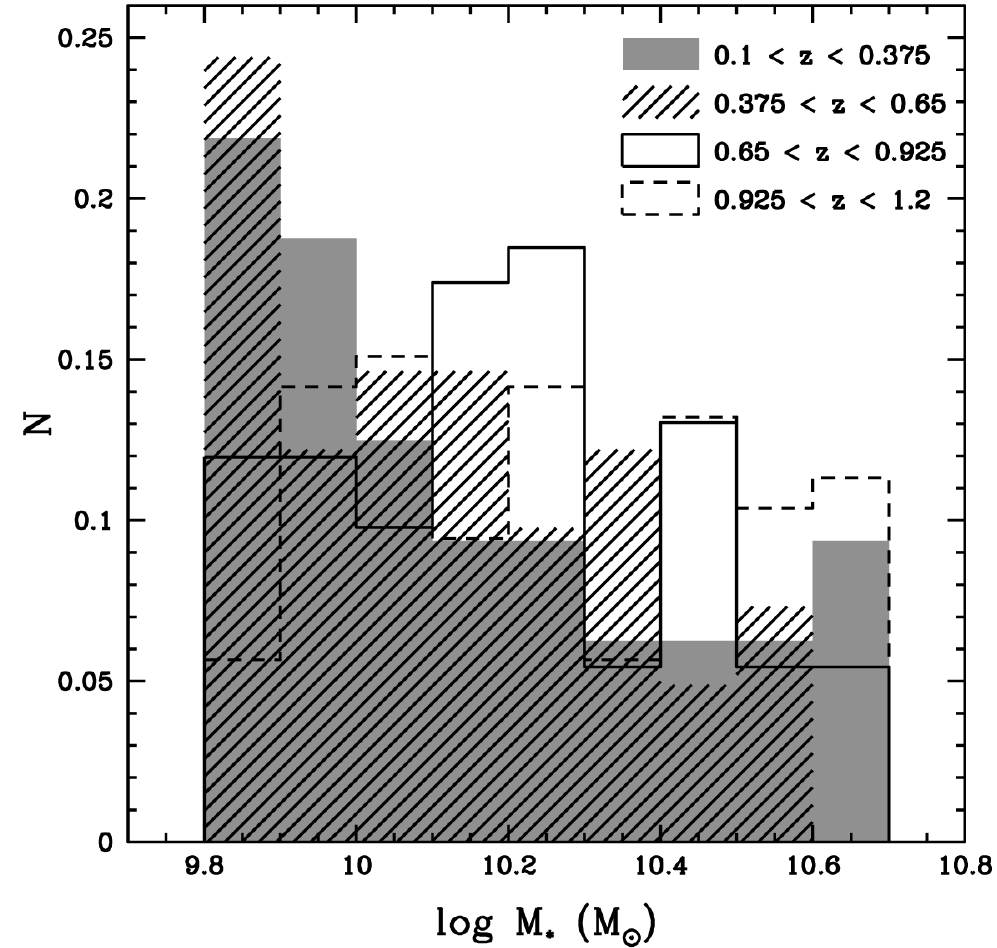}
\caption{For the mass-limited sample, the distribution of galaxy stellar masses among the redshift bins in Figure~\ref{velsig_mst} is shown.  It demonstrates 
that the average stellar mass of the sample is not the same at all redshifts: it is lower at lower redshift.  This is likely because higher mass
galaxies are leaving the sample to become early types.  This phenomenon works in the opposite direction to our results, making them
lower limits to the intrinsic evolution, as explained in \S5.2.
\label{mst_hist}}
\end{figure}

\section{Stellar Mass and Size Measurements}

In this paper we will study the relationship between galaxy kinematics and stellar mass, building upon the scaling relations studied in K07.
We adopt stellar mass measurements from \citet{lihwai} via private communication, as was done in K07.
The uncertainties on these stellar mass measurements are $\sim 0.25$ dex.

In addition to the effects of stellar mass on galaxy kinematics, we will also look at galaxy sizes.
Two size measurements are studied: emission line extent and continuum size.
In this paper we measure emission line extents from ground-based DEEP2 spectra from 
the same lines used to measure kinematics.  However, unlike the
kinematic measurements, no seeing correction is applied.
The spatial extent of the line emission is measured as follows for a given galaxy.  For details on the
method, see \citet{wei1}.  First we measure the spectral continuum by collapsing the two-dimensional 
spectrum in wavelength over a 15\AA\ range around the line and two 100\AA\ spatial ranges on either side of the line. 
We then fit a Gaussian profile along the slit using a non-linear least-squares routine and adopt the resulting
full-width-half-maximum (FWHM) as our measurement of emission line extent. 
For the galaxy continuum sizes, we adopt measurements made from $Hubble$ images 
using the {\tt GIM2D} software package \citep{gim2d} by \citealt{dutt} and E. Cheung et al. in preparation.
To take into account band-shifting with redshift, sizes are measured from $V$ and $I$-band 
images for galaxies at redshifts less than and greater than $z=0.6$, respectively.  
Since the continuum sizes are measured from $Hubble$ images, they are unaffected by seeing.

\section{Kinematic Measurements}

Measurements of internal galaxy kinematics from emission lines are adopted from K07.  In this 
section we discuss some relevant details and the phenomena they describe.  
Further details on how these measurements were made are given in \citet{wei1,wei2} for a similar survey  
with a lower spectral resolution: $R \sim 2100$ compared with $R \sim 5000$ for the K07 sample.  

For each galaxy the emission line used to measure kinematics is the one with the highest signal-to-noise 
in its spectrum.  For the vast majority of galaxies, the H$\alpha$ $\lambda 6563$,
[\ion{O}{2}] $\lambda\lambda 3726.0, 3728.8$, and  [\ion{O}{3}] $\lambda 5007$ lines are used.
Rotation velocities on the flat parts of the rotation curves ($V_{rot} \times sin(i)$), where $i$ is the inclination of a galaxy to the line-of-sight) 
and integrated gas velocity dispersions ($\sigma_g$) were measured simultaneously from the emission lines,
and the effects of seeing (typically $\sim 0.7$\arcsec) were taken into account.  
Our spectral resolution allows for $V_{rot} \times sin(i)$ and $\sigma_g$ to be measured 
down to $\sim 5$ and $\sim 15$ km s$^{-1}$, respectively.  
An uncertainty of 10 km s$^{-1}$ is adopted for both $V_{rot} \times sin(i)$ and $\sigma_g$ 
to account for random errors and the dependence of model parameters on the assumed seeing and scale radius
of the rotation curve.  Although many of the rotation curves do not
``turn over" due to seeing, our model is still able to fit for the rotation velocity on the flat part of the rotation curve \citep[][\S2.3.2]{wei1}.
In addition, it has been demonstrated using deeper data with the same telescope/instrument which observe a
turn-over in the rotation curve for 90\% of galaxies that spectra of our depth do not show a bias in 
$V_{rot}$ \citep[Figure 6 of][]{mill}.  Benefits of the deeper data \citep[as used in][]{mill} 
are smaller errors in $V_{rot} \times sin(i)$ (and therefore less scatter) and the ability to probe 
galaxies with fainter emission lines.  Except for the 24 galaxies for which reliable inclinations could not be determined (\S 2), 
values of $V_{rot} \times sin(i)$ are corrected for galaxy inclinations measured from the $Hubble$ images.
The results of this paper do not change significantly if these 24 galaxies are removed from the sample. 

\subsection{What $\sigma_g$ and $V_{rot}$ Measure}

{\it The integrated gas velocity dispersion we measure ($\sigma_g$) is unlike the typical
pressure-supported velocity dispersion measured for early type galaxies or galaxy bulges.}
This is the case for two reasons.  First, $\sigma_g$ is measured from emission lines
which trace the gas in galaxies, as opposed to absorption lines which trace stars.
Gas, unlike the collisionless stars in an early type galaxy, can dissipate energy and therefore 
cannot remain in a high dispersion equilibrium state with crossing orbits.
Secondly, because the spectral slits used in the DEEP2 Survey are 
wide compared to the apparent sizes of the galaxies observed ($1\arcsec$ slit vs. $\sim3\arcsec$ galaxies),
$\sigma_g$ effectively integrates velocity gradients
 on scales at and below the seeing limit \citep{wei1,covi,ghasp}.  For values of $\sigma_g$ less than $\sim35$ km s$^{-1}$, which is the upper limit for
well-ordered disk galaxies in the local Universe \citep[Figure 15 of][]{ghasp}, $\sigma_g$ measures the relative motions of individual star-forming regions in galaxy disks.  However, 
for values of $\sigma_g$ greater than $\sim35$ km s$^{-1}$, $\sigma_g$ integrates over velocity gradients which can correspond to non-ordered gas kinematics such as small-scale velocity gradients, gas motions due to star-formation, or superimposed clumps along the line of sight.  {\it In this paper, we collectively refer to these velocity gradients as ``disordered motions."}  

To test for interference from $V_{rot}$ in measurements of $\sigma_g$, in \S5 we compare $\sigma_g$ measured for galaxies in
our mass-limited sample with galaxies from DEEP2 in the same stellar mass range
which are face-on in $Hubble$ images (i.e., $i<30\degr$).
We find no significant difference in the trends of $\sigma_g$ with redshift
between these two samples, making it unlikely that the rotation velocity of a galaxy interferes with a measurement of its $\sigma_g$.

To further investigate what $\sigma_g$ and $V_{rot}$ measure, in \citet{covi} we performed mock observations
of a suite of simulations of major mergers of disk galaxies.  These simulations provided model galaxies in various stages of 
disorder, ranging the gamut from well-ordered disks to merger remnants.  The mock observations were performed 
just as the actual observations.  Galaxies were redshifted to $z\sim0.3$ and $z\sim1.0$
and seeing, slit width, and detector pixel size were taken into account.  
Our algorithm for fitting kinematics was found by \citet{covi} to be successful at reproducing $\sigma_g$, but 
was found to underestimate $V_{rot}$ by up to 30\%, independent of merger stage or redshift. 
However, as we note above, \citet{mill} did not find a systematic difference in $V_{rot}$ between the deeper data
and data similar to ours.  Therefore, if $V_{rot}$ is underestimated by up to 30\%, it is probably underestimated by this much 
in both our sample and in the deeper \citet{mill} dataset.

\begin{figure*}
\includegraphics[scale=0.9]{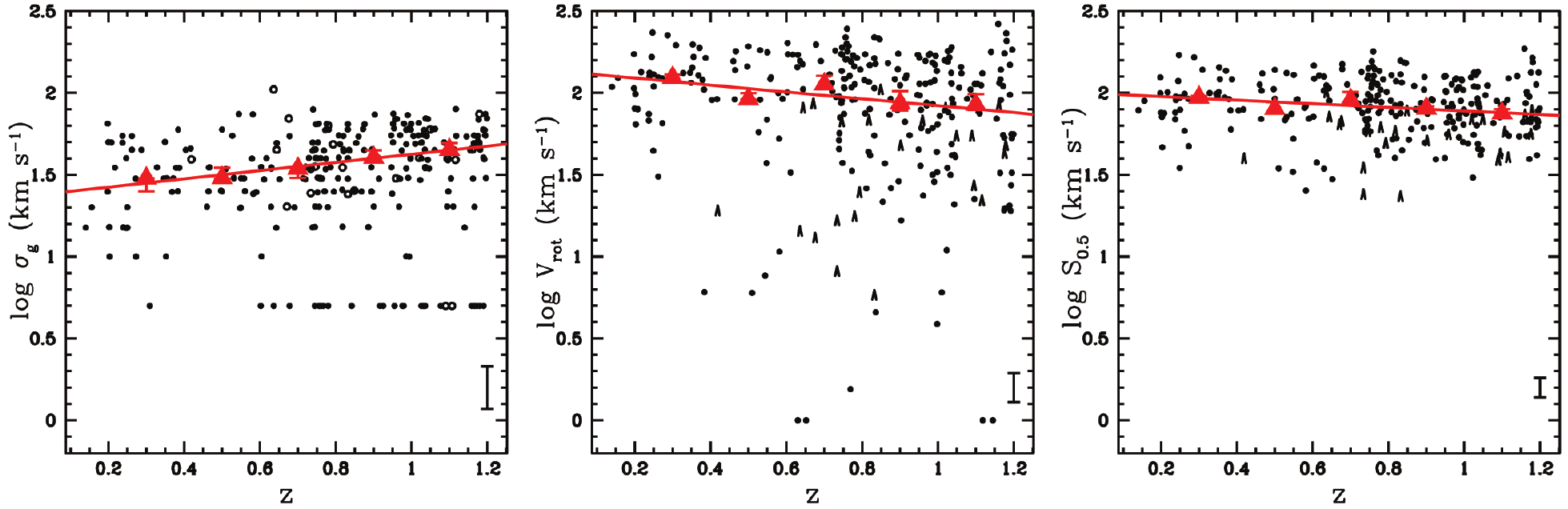}
\caption{The evolution of $\sigma_g,$ $V_{rot},$ and {\small $S_{0.5} \equiv \sqrt{0.5V_{rot}^2 + \sigma_g^2}$} with redshift for the mass-limited sample 
($9.8 < {\rm log}\,M_* (M_{\odot}) < 10.7$) are shown.  Galaxies increase in $\sigma_g$ and decrease in $V_{rot}$ and $S_{0.5}$ with increasing redshift.
Galaxies are plotted as black symbols, as described in Figure~\ref{tfr}.  Binned medians are shown as red triangles with error bars representing the error on the medians determined by bootstrap re-sampling the data.  Linear fits to the binned medians are shown as red lines.   Medians errors on the
individual measurements are shown in the lower right-hand corners of the plots, and include uncertainties in galaxy inclinations
for $V_{rot}$ and $S_{0.5}$.
\label{indiv}}
\end{figure*}

\begin{figure*}
\includegraphics[scale=0.9]{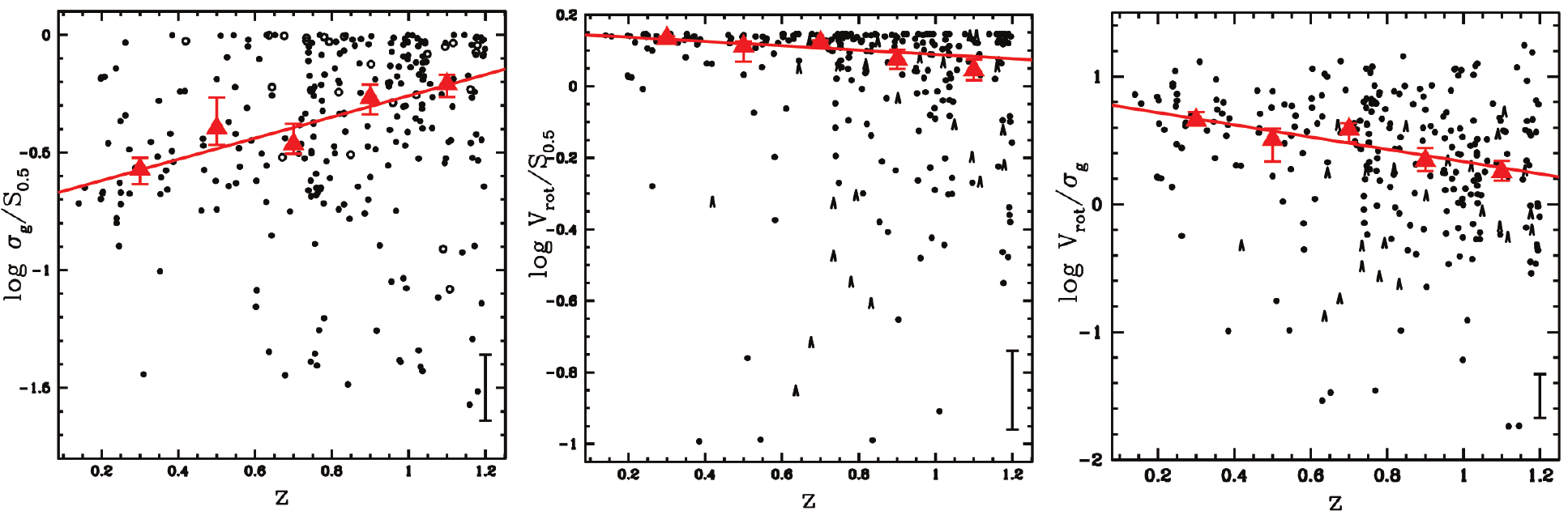}
\caption{The evolution of $\sigma_g/S_{0.5}$, $V_{rot}/S_{0.5}$, and $V_{rot}/\sigma_g$ with redshift for the mass-limited sample are shown.  The
first two of these ratios remove the dependence on potential well depth (i.e., $S_{0.5}$) in the $\sigma_g$ and $V_{rot}$ plots in Figure~\ref{indiv}.
Symbols, error bars, and fits are the same as in Figure~\ref{indiv}.
\label{ratios}}
\end{figure*}

\begin{figure}
\includegraphics[scale=0.88]{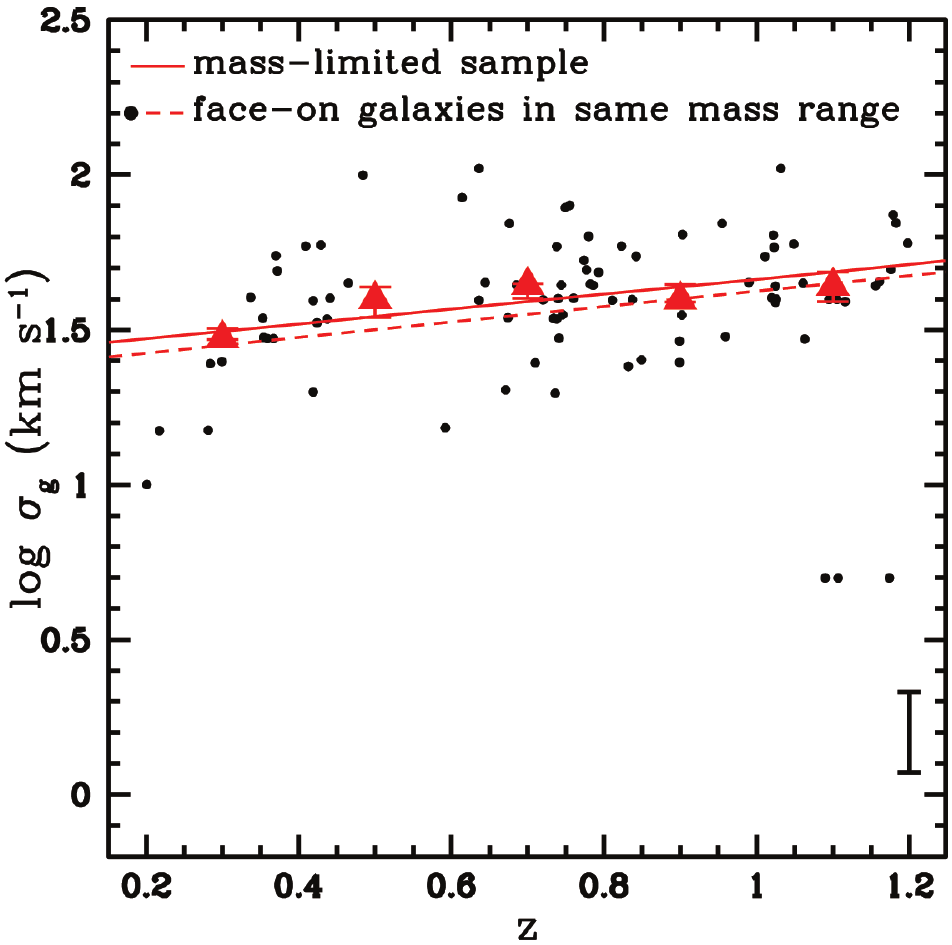}
\caption{The evolution of $\sigma_g$ with redshift for galaxies oriented face-on in $Hubble$ images ($i<30\degr$) 
(black points, red triangles show medians in redshift bins, solid red line shows fit to median points) is
compared with that for galaxies in the mass-limited sample from Figure~\ref{indiv} (dashed red line shows fit).  Face-on galaxies provide a 
pure measurement of $\sigma_g$, uncontaminated by $V_{rot}$.
Both samples span the same stellar mass range, and both fits are given in the text.
There is no significant difference between the fits, demonstrating that our measurements of $\sigma_g$ are unaffected by galaxy rotation.
\label{faceon}}
\end{figure}

\begin{figure*}[hb]
\includegraphics[scale=0.92]{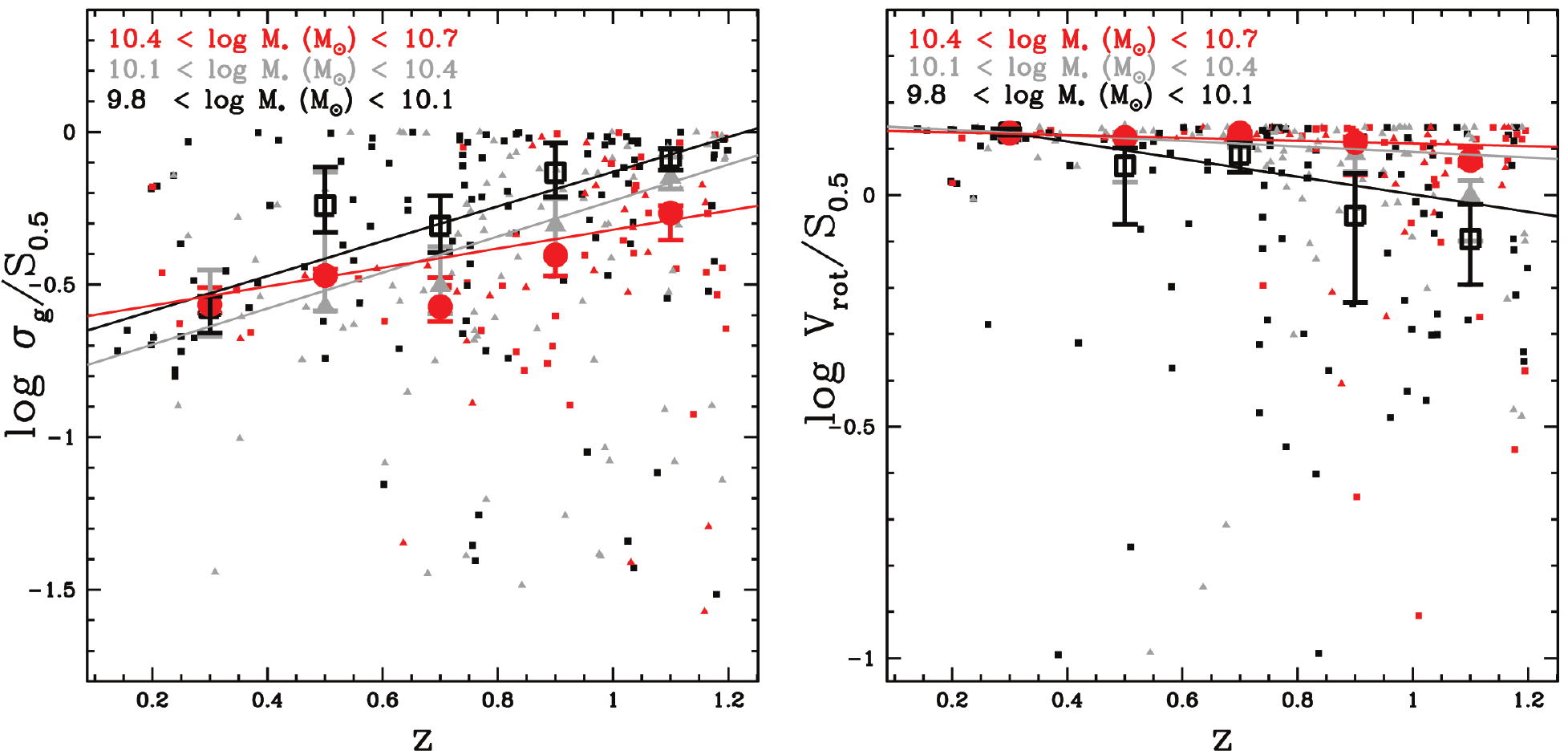}
\caption{As in Figure~\ref{ratios}, the quantities $\sigma_g/S_{0.5}$ and $V_{rot}/S_{0.5}$ are shown as a function of redshift for the mass-limited sample, but
here galaxies are divided into three mass bins: $9.8 < \rm log \ M_* (M_{\odot}) < 10.1$ (black squares),  $10.1 < \rm log \ M_* (M_{\odot}) < 10.4$ (grey triangles),
and $10.4 < \rm log \ M_* (M_{\odot}) < 10.7$ (red circles).  Individual galaxy points are small and medians points in redshift bins are enlarged. 
Medians are shown only for bins in redshift which have $>5$ galaxies.
All mass ranges show similar trends with redshift.  However, the higher-mass galaxies are consistently the 
most well-ordered at all redshifts (high $V_{rot}/S_{0.5}$ and low $\sigma_g/S_{0.5}$), the intermediate mass galaxies the next well-ordered, and the lowest 
mass galaxies the least well-ordered.   There appears to be a threshold at $\rm log \ M_* = 10.4 \ M_{\odot}$ separating galaxies with more ordered motions
from those with more disordered motions.  
\label{mass}}
\end{figure*}

\section{Evolution of Galaxy Kinematics}

In this section we examine how the kinematics of blue galaxies evolve with redshift.
First we examine the full K07 sample, and then we study a mass-limited sub-sample.
We return to the full sample again in \S7.

\subsection{The Full K07 Sample}

In Figure~\ref{tfr} we show $\sigma_g,$ $V_{rot},$ and $S_{0.5}$ versus stellar mass in bins of redshift
for the K07 sample.  The $V_{rot}$ and $S_{0.5}$ relations were already shown in K07; here we show the
$\sigma_g$ relation for completeness.  The $V_{rot}$  versus stellar mass relation (i.e., the stellar mass 
Tully-Fisher Relation) shows large scatter to low $V_{rot}$ at all redshifts (K07), as does $\sigma_g$ versus stellar mass.
However, when $V_{rot}$ and $\sigma_g$ are combined into $S_{0.5}$ (a tracer of potential well depth; \citealt{wei1}; K07), the resulting relation with stellar mass 
tightens significantly and does not evolve with redshift since $z=1.2$ to within uncertainties (K07).   
Figure~\ref{tfr} also shows that at lower redshift higher mass galaxies have on average larger values
of $V_{rot}$ and smaller values of $\sigma_g$ than at higher redshift (K07).  It is also clear from this figure that at lower redshift
lower mass galaxies have low values of $V_{rot}$ and high
values of $\sigma_g$ (K07); they are too faint to be observed at higher redshifts in our sample.

These trends are illustrated more clearly in Figure~\ref{velsig_mst} (left column) where the quantity $V_{rot}/\sigma_g$, the ratio of ordered to disordered  motions in a galaxy, is plotted versus stellar mass in bins of redshift.\footnote{In the evaluation of 
$V_{rot}/\sigma_g$, a minimum value of $\sigma_g$ of 15 km s$^{-1}$, which is the lowest value which we can measure, 
is adopted to avoid dividing by small numbers with relatively large uncertainties.}  The galaxies define an upper envelope 
in $V_{rot}/\sigma_g$ versus stellar mass-space
which does not evolve since $z=1.2$.  However, galaxies evolve within the envelope: with increasing time (i.e., decreasing redshift) 
they move away from the bottom of the plot toward the upper envelope, and this happens first for massive galaxies.    
This evolution is quantified by the histograms in Figure~\ref{velsig_mst} (right column) which demonstrate that galaxies
move towards higher $V_{rot}/\sigma_g$ with decreasing redshift (i.e., with time), and the highest-mass 
galaxies have on average the highest values of $V_{rot}/\sigma_g$ at all times.

\subsection{A Mass-Limited Galaxy Sample}

To investigate these trend of $V_{rot}/\sigma_g$ with redshift for all but the very lowest and very highest mass galaxies 
in our sample, we use a  mass-limited sample of galaxies with stellar masses over 
$9.8 < {\rm log}\,M_* (M_{\odot}) < 10.7$ (vertical dotted lines in Figures~\ref{colormass}-\ref{velsig_mst}),
as described in \S 2.
For the remainder of this section and for \S 6, we will focus on this mass-limited sample
of 270 galaxies.  In Figure~\ref{mst_hist}, for the mass-limited sample we plot the distribution of galaxy stellar masses
among the redshift bins in Figure~\ref{velsig_mst}.  It shows that the average stellar mass is not the same
for all redshift bins: at lower redshifts it is shifted towards lower masses. This is likely because higher 
mass galaxies are leaving the sample as they transform to red early
type galaxies \citep[e.g.,][]{fabe}.  This trend works in the opposite direction to our results, as we will
discuss later in this section. Therefore, the evolution we find is interpreted as a lower limit to the intrinsic evolution.

Figure~\ref{indiv} shows the evolution of $\sigma_g$, $V_{rot}$ and $S_{0.5}$ with redshift for the mass-limited sample.
All three quantities have large intrinsic scatter.  We will demonstrate that a portion of
the scatter is due to stellar mass or size in \S 6.  
Medians of the individual measurements in bins of redshift are shown as red triangles in Figure~\ref{indiv}.
Errors on these medians are calculated by bootstrap re-sampling the data in each redshift bin.
Bootstrap resampling is used as a standard method
for estimating the statistical error without assumptions about the shape of the distribution.
We re-sampled with replacement from the data 1000 times, and calculated the dispersion of the 
sample medians.  This dispersion is adopted as the error on the median points.  It accounts
for the spread in the individual data points due to both intrinsic variation and observational error.
The errors on the medians are small because  the sample is large and there are many points in each
redshift bin.  

We fit linear relations to $\sigma_g$, $V_{rot}$ and $S_{0.5}$ versus redshift by 
performing linear least-squares fits to the binned medians, taking into account the errors on the medians.
The results are not sensitive to the redshift bins used.  We fit to the medians rather than directly to the data
because the distributions of the data points are non-Gaussian
and the intrinsic spread of the points is significant compared to the
individual error bars, which would cause standard least squares fits to be
misleading.  
To avoid covariances on the fitted parameters, the medians are zero-pointed near the middle of the 
sample when performing the fits.  Of the three quantities, the median $\sigma_g$ shows the
strongest evolution ($5.0 \sigma$).  It increases linearly with redshift to $z=1.2$ (i.e., decreases with time) as  
\begin{equation}
{\rm log}\ \sigma_g - 0.42 = (0.25 \pm 0.05) (z - 0.82) + (1.16 \pm 0.01),
\end{equation}
and the fit has a $\chi^2$  od 1.1.  The median $\sigma_g$ increases from $27\pm1$ km s$^{-1}$ at $z=0.2$ to $47\pm1$ km s$^{-1}$ at $z=1.2$.
The quantity which shows the second-strongest evolution is the median $V_{rot}$ ($4.2 \sigma$).  It decreases with redshift as
\begin{equation}
{\rm log}\ V_{rot} - 2.00 = (-0.21 \pm 0.05) (z - 0.82) - (0.04 \pm 0.02),
\end{equation}
and the fits has a $\chi^2$ of 16.1.  The median $V_{rot}$ decreases from $123\pm1$ km s$^{-1}$ at 
$z=0.2$ to $76\pm1$ km s$^{-1}$ at $z=1.2$.   Finally, the median $S_{0.5}$ shows the weakest
evolution of the three kinematic quantities ($3.6 \sigma$).  It decreases with redshift as
\begin{equation}
{\rm log}\ S_{0.5} - 1.92 = (-0.11 \pm 0.04) (z - 0.82) + (0.00 \pm 0.01),
\end{equation}
and the fit has a $\chi^2$ of 2.8.  The median $S_{0.5}$ decreases from $97\pm1$ km s$^{-1}$ at $z=0.2$ to $75\pm1$ km s$^{-1}$ 
at $z=1.2$.   

Some of the evolution found may be due to a changing distribution of
galaxy masses in the sample with time.  This can be caused by galaxies growing in mass
and/or high mass galaxies quenching their star-formation and leaving the sample.
To address this we remove the dependence on potential well depth 
by examining ratios with $S_{0.5}$, namely $V_{rot}/S_{0.5}$ and $\sigma_g/S_{0.5}$.
These ratios are shown as a function of redshift in Figure~\ref{ratios}.  They 
express the fraction of kinematic support which comes from ordered and disordered motions.
The median of the ratio $V_{rot}/S_{0.5}$ decreases with redshift ($3.0 \sigma$) as
\begin{equation}
{\rm log}\ V_{rot}/S_{0.5}  - 0.10 = (-0.06 \pm 0.02) (z - 0.82) + (0.00 \pm 0.01),
\end{equation}
and the fit has a $\chi^2$ of 4.5.  The median of the ratio $\sigma_g/S_{0.5}$ increases with redshift ($5.0 \sigma$) as
\begin{equation}
{\rm log}\ \sigma_g/S_{0.5}  + 0.35 = (0.45 \pm 0.09) (z - 0.82) + (0.01 \pm 0.03),
\end{equation}
and the fit has a $\chi^2$ of 2.3.
As before, these are linear least-squares fits to the median points.  
These ratios demonstrate that the increasing role of $\sigma_g$ and the declining
role of $V_{rot}$ with increasing redshift are not due to a changing distribution of galaxy masses in the sample.

The evolution found is interpreted as a lower limit to the intrinsic evolution.
As discussed earlier in this section and shown in Figure~\ref{mst_hist}, the average stellar mass of our mass-limited sample
is lower at lower redshifts.  From Figure~\ref{velsig_mst} it is apparent that on average low mass 
galaxies have higher $\sigma_g$ and lower $V_{rot}$ than more massive galaxies at any redshift (we will investigate this
further in \S6).  Therefore, our findings that for galaxies in the mass-limited sample the average $\sigma_g$ and $V_{rot}$ are lower and higher, respectively, at
lower redshift than at higher redshift are interpreted as lower limits to the intrinsic evolution.

In addition to ratios with $S_{0.5}$, we look at $V_{rot}/\sigma_g$, a ratio of ordered to disordered motions$^{13}$.
This is not an ideal quantity since neither $V_{rot}$ nor $\sigma_g$ is
constant with redshift.  Nevertheless, due to the significant trends of $\sigma_g$ and $V_{rot}$ with
redshift, the median $V_{rot}/\sigma_g$ also shows strong evolution ($4.8 \sigma$; Figure~\ref{ratios}):
\begin{equation}
{\rm log}\ V_{rot}/\sigma_g  - 1.55 = (-0.48 \pm 0.10) (z - 0.82) - (1.13 \pm 0.04)
\end{equation}
with $\chi^2=2.8$.

To determine whether the $V_{rot}$ of a galaxy influences the measurement of $\sigma_g$, in
Figure~\ref{faceon} we examine the evolution of $\sigma_g$ with redshift for galaxies
which are face-on in $Hubble$ images (i.e., $i \le 30$\degr) and in the same stellar mass
range as the mass-limited sample.  This is a subset of galaxies for which $\sigma_g$ should be
completely secure.  We find the same evolution in the median $\sigma_g$ to within uncertainties for face-on galaxies: 
$log\ \sigma_g - 0.42 = (0.24 \pm 0.04) (z - 0.82) + (1.20 \pm 0.02)$, demonstrating that $V_{rot}$ does not
influence the measurement of $\sigma_g$.

\section{Downsizing}

In this section we demonstrate that much of the scatter in the relations of $\sigma_g/S_{0.5}$ and $V_{rot}/S_{0.5}$ with redshift
is attributable to galaxy properties such as stellar mass and size, and find that trends with stellar mass and size are of similar strengths.
We also demonstrate that there is a threshold in stellar mass and size acting at all redshifts which separates galaxies with high $\sigma_g/S_{0.5}$ 
and low $V_{rot}/S_{0.5}$ from those with low $\sigma_g/S_{0.5}$ and high $V_{rot}/S_{0.5}$.

First we consider stellar mass.  In Figure~\ref{mass}, as in Figure~\ref{ratios}, we show $\sigma_g/S_{0.5}$ and $V_{rot}/S_{0.5}$
as a function of redshift for the mass-limited sample.  However, in Figure~\ref{mass} galaxies are  divided into three stellar mass bins:
$9.8 < {\rm log}\,M_* (M_{\odot}) < 10.1$,  $10.1 < {\rm log}\,M_* (M_{\odot}) < 10.4$, and $10.4 < {\rm log}\,M_* (M_{\odot}) < 10.7$.
Similar trends to those found for the entire mass-limited sample 
are also found separately for the three mass bins. Interestingly, it is also evident 
that at all redshifts examined higher mass galaxies are the most kinematically settled (i.e., higher $V_{rot}/S_{0.5}$ and lower $\sigma_g/S_{0.5}$),
intermediate mass galaxies the next most settled, and low mass galaxies the least settled.
This is consistent with downsizing trends in which
more massive galaxies attain final structure and star-formation rates sooner \citep[e.g.,][]{cowie}, and
we refer to this phenomenon as ``kinematic downsizing."  Furthermore, there appears to be a threshold
at log $M_* = 10.4 \ M_{\odot}$ which separates galaxies with high $\sigma_g/S_{0.5}$ and low $V_{rot}/S_{0.5}$
from those with log $\sigma_g/S_{0.5}$ and high $V_{rot}/S_{0.5}$.  This is most evident in the median points
in Figure~\ref{mass}: those for the lower and intermediate mass bins are very similar, while those for the higher mass bin differ significantly
from the other mass bins.

\begin{figure*}[p]
\includegraphics[scale=0.91]{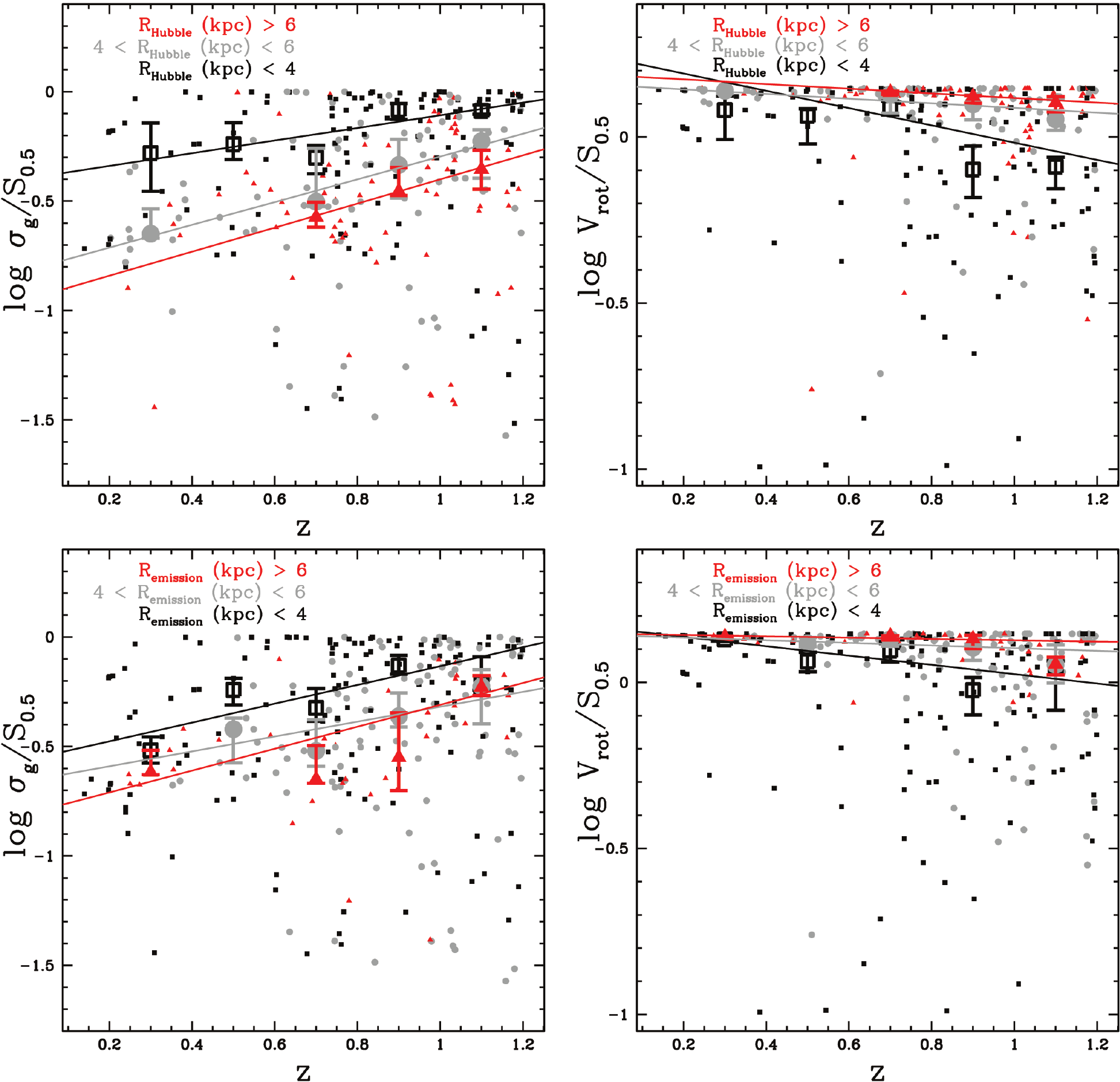}
\caption{This figure is the same as Figure~\ref{mass}, except galaxies from the mass-limited sample are divided into three size bins: 
$<4$ kpc (black squares), 4--6 kpc (grey circles), and $>6$ kpc (red triangles).   
Two size measures are considered: $Hubble$ half-light radius which is not affected by seeing (top two plots)  and emission line extent which is affected
by seeing and not corrected for its influence (bottom two plots).  Galaxies in size bins show similar trends to the full
mass-limited sample.  However, at all redshifts the largest galaxies are the most well-ordered (high $V_{rot}/S_{0.5}$, 
low $\sigma_g/S_{0.5}$), the intermediate-sized galaxies the next well-ordered, and the smallest galaxies the least well-ordered.  
There appears to be a threshold at 4 kpc which corresponds to the threshold in stellar mass seen in Figure~\ref{mass}.
\label{size}}
\end{figure*}

Next we examine trends with size.  We consider two size measurements: $Hubble$ continuum sizes 
and emission line extents, as described in \S 3.  $Hubble$ sizes are unaffected by seeing, and
emission line extents are affected by seeing and are {\it not} corrected for it.
Figure~\ref{size} is similar to Figure~\ref{mass} except
galaxies are color-coded according to size instead of stellar mass.  For both size measurements, three size bins are
shown: $<4$ kpc, 4--6 kpc, and $>6$ kpc.  Figure~\ref{size} shows that on average at all redshifts examined
the largest galaxies are the most kinematically well-ordered, the smallest the most disordered, and
intermediate-sized galaxies have properties in between the two.
This is the case for both size measurements.  It demonstrates that emission lines,
which we use to measure galaxy kinematics, are likely good tracers of the spatial distribution of the galaxy continua
observed in $Hubble$ images.  It also demonstrates that we
are able to measure reliable rotation velocities for even the small galaxies in our sample.
Similar to the stellar mass bins, there is also a size threshold.  It is at 4.0 kpc, which corresponds to 
the threshold in stellar mass (log $M_* = 10.4 \ M_{\odot}$) in the size versus stellar mass relation for the mass-limited sample.

Figures~\ref{mass} and \ref{size} also demonstrate that the evolution in Figures~\ref{indiv} and \ref{ratios} 
is not due to a special population of high-redshift galaxies with high values of $\sigma_g$ and low values of 
$V_{rot}$.  It occurs even if only massive/large galaxies are considered.  

\clearpage

\section{Settling Fraction}

\begin{figure*}[!h]
\includegraphics[scale=3.5]{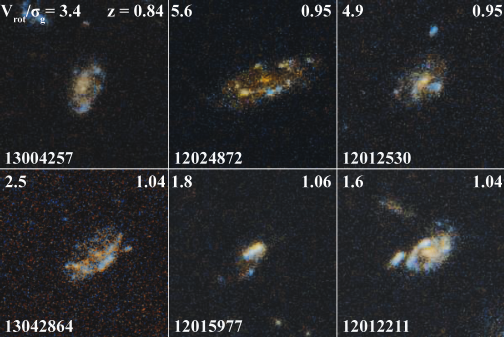}
\caption{$Hubble/ACS$ $V+I$ images of settled (top) and unsettled (bottom) galaxies which are defined
as having $V_{rot}/\sigma_g >3$ and $V_{rot}/\sigma_g < 3$, respectively.  The settled galaxies have normal 
disk-like morphologies and the unsettled galaxies have disturbed morphologies, as determined
via eyeball inspection (K07; Kassin et al.\,in preparation).  All galaxies have stellar masses 
in the range $10.2 < {\rm log}\,M_* (M_{\odot}) < 11.1$, 
and the images are 6\arcsec\ on a side.  
\label{morphs}}
\end{figure*}

To better quantify how blue galaxies evolve in $\sigma_g$ and $V_{rot}$, we use these parameters to define a
kinematically settled disk galaxy and then study the fraction of settled disk galaxies as a function of stellar mass and redshift.
We return again to the full K07 galaxy sample.  For the definition of a settled disk, we first look to the distribution 
of $V_{rot}/\sigma_g$ in Figure~\ref{velsig_mst} and then to $Hubble$ morphologies.  Since we regard higher-mass 
blue galaxies to be on average kinematically settled disks at low redshift, we look to their values of $V_{rot}/\sigma_g$ in Figure~\ref{velsig_mst}.  A typical value for galaxies with log $M_* (M_{\odot}) > 9.8$ is $\sim3$ (i.e., log $V_{rot}/\sigma_g = 0.48$).  We now consider  
$Hubble$ morphologies.  K07 and S. Kassin et al.\,(in preparation) show that for the full K07 sample, $Hubble$ morphologies 
are correlated with $V_{rot}/\sigma_g$ such that galaxies with normal disk-like morphologies via eyeball inspection have higher
values than galaxies which appear more disturbed.  We find that $V_{rot}/\sigma_g \sim 3$ indeed provides a 
reasonable division between normal disk-like and disturbed morphologies,   
as demonstrated by $Hubble$ images of 6 galaxies from the K07 sample in Figure~\ref{morphs}.  
In addition, $V_{rot}/\sigma_g \sim 3$ provides a good separation between the ``rotating disks" and 
``perturbed rotators" in the IMAGES Survey at $z\sim0.6$ \citep[][Figure 6]{puec_j}.
The median $V_{rot}/\sigma_g$ for galaxies in the local GHASP Survey \citep{ghasp} is 6.3 with 
an rms scatter of 3.3, indicating that disk galaxies in the local universe are even more kinematically
quiet than those which pass our ``settled" criterion.

\begin{figure}
\includegraphics[scale=0.85]{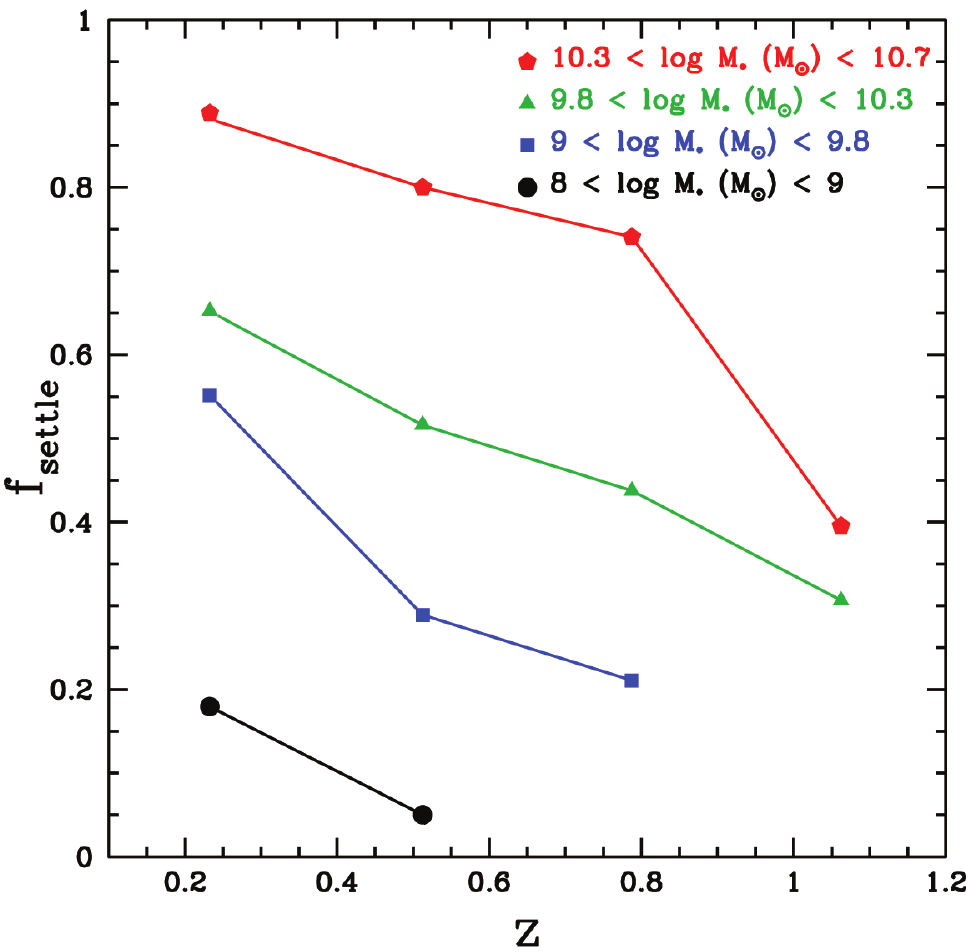}
\caption{The fraction of galaxies in the full K07 sample which are settled (defined as $V_{rot}/\sigma_g>3$), $f_{settle}$,
is shown as a function of redshift for four  bins in stellar mass.  The quantity $f_{settle}$ decreases with redshift (increases with time) 
for all mass bins.  At all redshifts the higher the stellar mass of the population,
the higher its $f_{settle}$.   
\label{settle}}
\end{figure}

This quantitative definition of a settled disk galaxy (i.e., $V_{rot}/\sigma_g > 3$) can be used to measure
the fraction of blue galaxies which are settled, $f_{settle}$.  The evolution of $f_{settle}$ with redshift for 
the full K07 sample divided into in four stellar mass bins is shown in Figure~\ref{settle}.
All bins show increasing $f_{settle}$ with decreasing redshift (i.e., increasing
$f_{settle}$ with time).  For example, $f_{settle}$ increases over $0.2<z<1.2$ from 40\% to 89\% for $10.3 < {\rm log}\,M_* (M_{\odot}) < 10.7$
and from 31\% to 65\% for $9.8 < {\rm log}\,M_* (M_{\odot}) < 10.3$.  For the less massive bins,
we probe smaller ranges in redshift.  These are lower limits to the intrinsic evolution, as discussed in \S5.2.
Furthermore, at all redshifts the higher the stellar mass of the galaxy population, the higher its value of $f_{settle}$.
The same qualitative behavior is found for  settled galaxies defined as  $V_{rot}/\sigma_g > 1-4$,
except for the lowest mass bin.

\section{Kinematic Measurements in the Literature}

\begin{figure*}
\begin{center}
\includegraphics[scale=0.77]{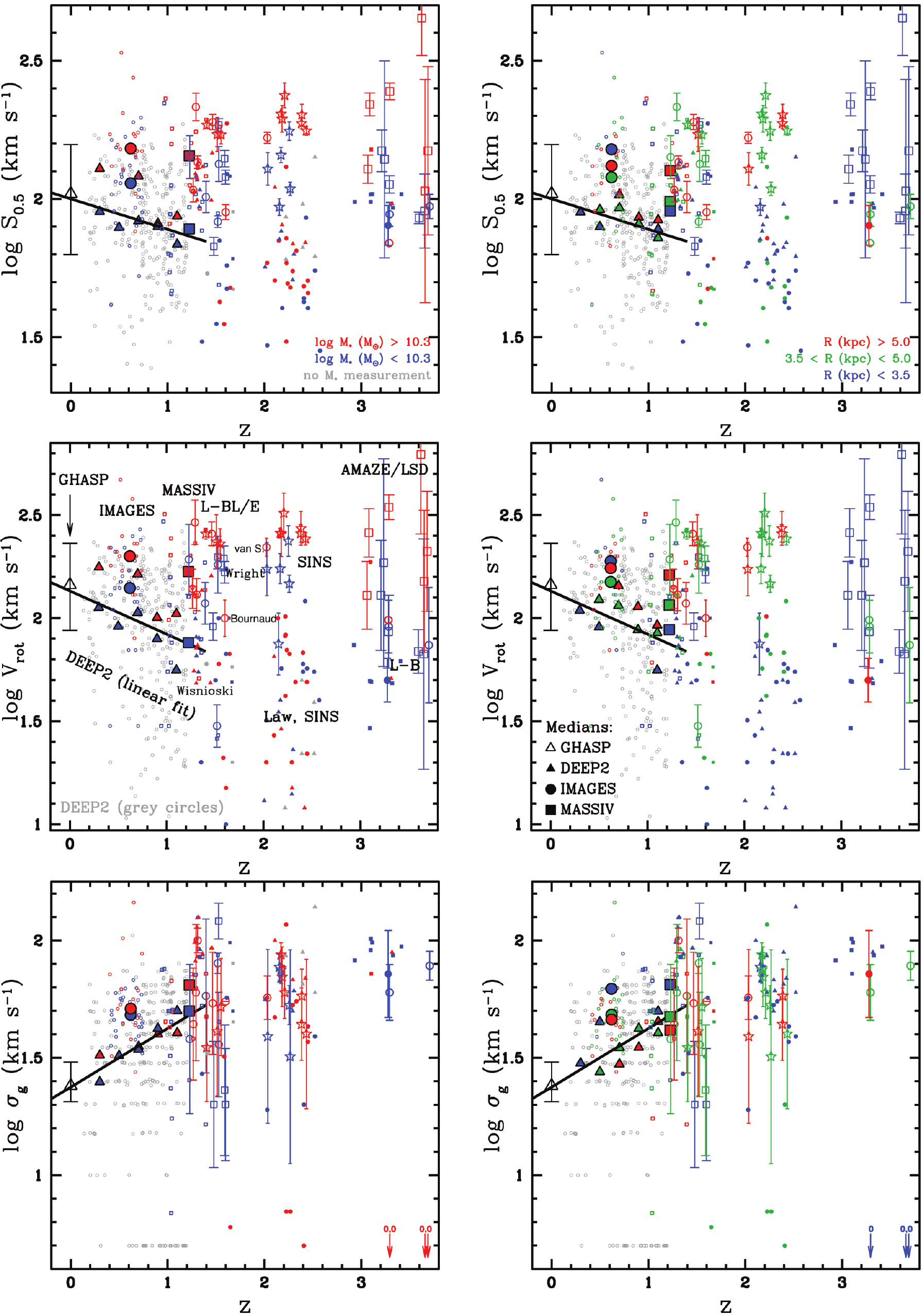}
\end{center}
\caption{The quantities plotted here are the same those shown in Figure~\ref{indiv}, except here the $z$ axes are expanded to include 
data from the literature.  In the plots on the left and right galaxies are 
color-coded according to stellar mass and emission line extent, respectively, unless noted in \S 8.  For the local 
GHASP survey, only sample medians are shown (large open black triangles) with error bars depicting the rms scatter.  
Large filled symbols denote medians of large samples at $z\la1.2$: the mass-limited sample from this paper (``DEEP2;" triangles),
IMAGES (circles), and MASSIV (squares).  Individual galaxies from these large samples are shown as
small open symbols: DEEP2 (grey circles; not colored according to mass/size), IMAGES (circles), 
and MASSIV (squares).  At $z\ga1.2$, only individual galaxies are shown: large open symbols with error bars denote 
galaxies where rotating disk models were fit, and small filled symbols without error bars denote galaxies where models were not fit
(for these galaxies, $V_{shear}$  and an average $\sigma_g$ are plotted).   Galaxies from \citealt{wisi} are shown as small filled triangles,
from \citealt{wright7,wright9} as large open squares, from \citealt{lem15} and \citealt{epin9} (``L-BL/E") as large open circles, 
from \citealt{bour08} as a large open circle, from \citealt{vans} as a large open circle,
from SINS as large open stars and small filled circles, from \citealt{law} as small filled triangles, from
\citet{lem3} (``L-BL") as large open circles and a single filled circle, and from AMAZE/LSD as large open squares and small filled squares.
\label{lit}}
\end{figure*}

In this section we compare our mass-limited sample with measurements in the literature for local galaxies and for galaxies at similar redshifts.
We also compile kinematic measurements from the literature and for galaxies at higher redshifts.
Measurements of $S_{0.5}, V_{rot}$, and $\sigma_g$ from the literature are shown in Figure~\ref{lit}, and 
are detailed below in order of increasing redshift.  
In the panels on the left and right in Figure~\ref{lit}, galaxies are color-coded according to stellar mass
(for a \citealt{chab} initial mass function) and emission-line extent, respectively, unless noted below.
The mass-limited sample is also shown in Figure~\ref{lit}
as small grey open circles (labeled ``DEEP2") with sample medians
shown as large filled triangles and linear fits to the medians as solid black lines.  
All literature measurements are from integral field spectroscopy (IFS).

To compare our dataet with local galaxies, a large homogenous sample 
is needed, and the GHASP survey \citep{ghasp} provides this.  Figure~\ref{lit}
shows median values of $S_{0.5}, V_{rot}$, and $\sigma_g$ for galaxies in GHASP as large open black triangles at $z=0$.
Error bars on these points depict the rms scatter of the sample.
Although GHASP spans a larger range in stellar mass  than our mass-limited sample:
$9.0 < {\rm log}\,M_* (M_{\odot}) < 11.7$ for a `diet' Salpeter IMF \citep{epin8} (which 
is roughly 0.15 dex higher than the \citealt{chab} IMF used in this paper),
the sample medians are consistent with extrapolations of our relations to $z=0$.
Individual stellar mass measurements for galaxies in GHASP are unavailable to create a more comparable sample.

Kinematic measurements for 47 out of 68 galaxies in the IMAGES IFS Survey which 
have stellar mass measurements in the same range as our mass-limited sample
are shown as small open circles in Figure~\ref{lit} 
(Puech private communication, \citealt{puec_images3}, \citealt{neic}, \citealt{yang_images1}).  
Sample medians are shown as large filled circles at the median redshift of the survey, $z=0.6$.
The sizes used are continuum half-light radii.
Although IMAGES has a similar selection to our survey, the median $V_{rot}$ and $\sigma_g$ of IMAGES are
larger than ours at $z=0.6$ by 0.23 and 0.18 dex, respectively.
The measurement methods are sufficiently different for IMAGES compared to our
methods that it is difficult to compare the two datasets.  However,
it is possible that the IMAGES median $\sigma_g$ is higher because, unlike our procedure of fitting
for $V_{rot}$ and $\sigma_g$ simultaneously, they are fit for separately in IMAGES.  
This may result in $\sigma_g$ measurements which are biased high because pixels with the highest 
signal-to-noise will be ignored when they are fit for separately \citep[e.g.,][]{davi}.  


The remaining points in Figure~\ref{lit} plot data for galaxies beyond $z=1.2$.  Kinematic data are difficult to obtain
at these redshifts since emission lines are redshifted to
near-infrared wavelengths where the sky is bright in continuum and line emission, making spectroscopy difficult.
Pioneering studies of galaxy kinematics at $z\ga1$ 
with single-slit near-infrared spectrographs gave the first glimpse of the kinematic state 
of galaxies at these earlier times, albeit only the most highly star-forming systems \citep[e.g.,][]{erb04, erb06}.    
More recently, there has been a boom in IFS observations of galaxy kinematics at these redshifts.
However, to obtain enough signal-to-noise to measure kinematics in a reasonable amount of observing time, 
all samples at these redshifts are unfortunately still biased toward galaxies with the brightest emission lines, which are the most highly star-forming systems.

The first set of points beyond $z\sim1.2$ in Figure~\ref{lit} shows kinematic 
measurements for 29 out of 46 galaxies from the MASSIV
Survey \citep[small open squares;][]{cont, verg}.   These galaxies have stellar masses in the same range
as our mass-limited sample and a median redshift of 1.2.  Medians in $S_{0.5}$, $V_{rot}$, and $\sigma_g$
for the MASSIV sample are shown as large filled squares.
The lower mass galaxies in MASSIV ($9.8 < {\rm log}\,M_* (M_{\odot}) < 10.3$) have median values of 
these quantities which are consistent with the lower mass galaxies in our mass-limited sample.  However, the higher mass galaxies
($10.3 < {\rm log}\,M_* (M_{\odot}) < 10.7$) have median values which are greater than the higher mass
galaxies in our mass-limited sample.  This is especially the case for $V_{rot}$ which has a median value which is 0.20 dex larger than
that for the higher mass galaxies in our mass-limited sample at $z=1$.  This difference is most likely
due to sample selection, since galaxies in MASSIV are chosen to be the very brightest galaxies in the VVDS Survey \citep{cont}.

Moving to higher redshift in Figure~\ref{lit}, we note that
many IFS observations of galaxies at $z \ga 1.5$ have kinematics which cannot be fit by a rotating disk model.
This is due to a few factors: a non-ordered velocity field, small spatial extent, and/or low signal-to-noise.  For these galaxies
we adopt half of the velocity shear across the face of the system ($\onehalf \times V_{shear}$) as $V_{rot}$,
where $V_{shear}$ is the difference between the minimum and maximum
rotation velocities and is not corrected for inclination.  These galaxies are shown as {\it small filled symbols without error bars} in Figure~\ref{lit}
to differentiate them from galaxies which were modeled as rotating disks and are shown as {\it large open symbols with error bars}.

For 13 extremely high star-forming galaxies at $z\sim1.3$, we adopt measurements from \citet{wisi}
which are not modeled as rotating disks (small filled triangles in Figure~\ref{lit}).
Next, for 8 galaxies at $z\sim1.5$, we adopt measurements from S. Wright (private communication) for 
observations presented in \citet{wright7,wright9}.  Galaxies which consist of two or more 
spatially separated  components are treated as individual systems.  Four of these galaxies were modeled as
rotating disks (large open squares) and four were not (small filled squares).  For galaxies which were so modeled,
values of $V_{rot}$ are {\it not} corrected for galaxy inclinations.    For the \citet{wright7,wright9} sample, all but the highest value of
$\sigma_g$ can be explained by a residual  seeing halo, and are therefore not likely physical.
For an additional 8 galaxies at $z\sim1.5$, we adopt data from \citet{epin9} and \citet{lem15} which analyze 
the same observations.   We abbreviate these references as ``L-BL/E."
For both studies, 8 galaxies are modeled as rotating disks.  Values of $V_{rot}$ 
and $\sigma_g$ are consistent between the two studies to within uncertainties except for one galaxy, VVDS 020147106.
For this galaxy we adopt $V_{rot}$ from \citet{epin9} since their model does not
extrapolate significantly beyond the observed velocity field.  For the remaining 7 galaxies, fitted quantities from \citet{lem15} 
are adopted.  These 8 galaxies are shown in Figure~\ref{lit} as large open circles.   We do not show data for two additional
galaxies from \citet{lem15} which were not able to be modeled, one of which is also in \citet{epin9}.
One of these galaxies, VVDS 020461235, does not have a measurement of $\sigma_g$.  The other galaxy, 
VVDS 020116027, has a velocity field which differs significantly between
the two studies, likely indicating low signal-to-noise.  Finally, we show best-fit parameters 
for one galaxy at $z=1.6$ from \citet{bour08} as large open circles in Figure~\ref{lit}.

For galaxies at $z\sim2$, data are adopted from the SINS Survey via private communication from N. F\"{o}rster-Schreiber for
the most recent models and data as of August 2011 for galaxies in \citet{genz}, \citet{fors9}, and \citet{cres}.
A total of 14 SINS galaxies were modeled as rotating disks (large open stars in Figure~\ref{lit}), and 33 were not (small filled circles).
For galaxies from SINS which were not modeled, we
estimate $\sigma_g$  as the line-width less the velocity shear (i.e., $v_{obs}$
from \citealt{fors9} less $V_{shear}$).  For an additional 16 galaxies at $z\sim2$ we adopt data from \citet{law}.
None of these galaxies were modeled as rotating disks, and they are shown as small filled triangles in Figure~\ref{lit}.  Three galaxies from \citet{law}
consist of two spatially separated components; we consider each of these components as a separate galaxy.
In addition, best-fit parameters for one galaxy at $z=2.03$ from \citet{vans} are shown as large open circles labeled ``van S."

For 4 galaxies at $z\sim3$, data are adopted from \citet{lem3}.  Three of 
these galaxies are modeled as rotating disks (large open circles in Figure~\ref{lit} labeled ``L-B") 
and one is not (filled circle shown enlarged in Figure~\ref{lit} because it overlaps an error bar for a different galaxy in the $S_{0.5}$ and $\sigma_g$  plots). 
In addition, measurements for 19 galaxies at $z\sim3$ data are adopted from the AMAZE/LSD Survey via private communication from A. Gnerucci for
data from \citet{gner} and Troncoso et al.\,(in preparation).  A total of
12 of these galaxies were modeled as rotating disks (large open squares) and 7 were not (small filled squares).

We will now attempt to summarize the disparate datasets from the literature over $1.2 < z <2.5$.  
Overall, there is large scatter in $\sigma_g$, $V_{rot}$, and $S_{0.5}$, and no clear trends with redshift.
While some of the scatter is likely due to the range in stellar masses probed, that
there are no trends is probably also due to other factors.  Namely,
only the brightest galaxies are observed at these redshifts, and 
instrumental effects and poor observational conditions likely play roles.
It is also clear that all high redshift galaxies observed have significant contributions to their kinematics from $\sigma_g$.

On average, distant galaxies beyond the last redshift of our sample divide into two populations \citep[e.g.,][]{law,fors9}: 
(1) The first consists of large massive galaxies with $V_{rot}$ 
equal to or slightly greater than the largest values which we measure for
DEEP2 galaxies, and which have significant values of $\sigma_g$. 
(2) The second population consists of small lower mass
galaxies with low or negligible $V_{rot}$ but with significant $\sigma_g$.  
These small galaxies are prominent in the \citet{law} sample and the SINS survey, but
are also found at lower redshifts in the DEEP2, IMAGES and MASSIV surveys.
Some of the observations at $z>1.2$ may be too shallow to detect the full rotation gradients, so
until these galaxies can be observed with significantly longer exposure times to detect possible faint disks, whether
or not all of them lack significant $V_{rot}$ is unclear.  Interestingly,
at $z \simeq 3$, nearly all galaxies are small but have large values of $V_{rot}$ and $\sigma_g$.

\section{Conclusions}

We study the internal kinematics of 544 blue galaxies over the last $\sim 8$ billion years,
and find significant evolution.  Blue galaxies become progressively more well-ordered with time as disordered motions decrease
and rotation velocities increase.   In addition, galaxy potential well depths continuously increase with time.  
At all redshifts the most massive galaxies are on average the most kinematically settled, and the least massive galaxies
the least kinematically settled.
The lowest mass galaxies in our survey, which are observed only at low redshift, are not well-ordered even today.
They may be in the process of settling, or may never become settled disks.  There appears to be a threshold at 
log $M_* = 10.4 \ M_{\odot}$ and 4.0 kpc which separates galaxies with more ordered kinematics from those with more disordered kinematics.
All in all, these trends with mass are consistent  with downsizing trends for other galaxy properties, and we refer to 
them as ``kinematic downsizing."

We define a kinematically settled disk galaxy as having a ratio of ordered to disordered motions ($V_{rot}/\sigma_g$)
greater than 3.  We find that the fraction of settled disk galaxies, $f_{settle}$, increases with time since $z=1.2$
for all galaxies over $8.0 < {\rm log \ M_* (M_{\odot})} < 10.7$.
The fraction $f_{settle}$ is larger for more massive galaxy populations, independent of redshift.  
These qualitative findings do not change if a settled disk galaxy is defined with $V_{rot}/\sigma_g > 1-4$. 

We speculate that the steady kinematic settling seen in our data is due to a combination of factors:
(1.) The frequency of mergers may be decreasing with time.  Merging, major but more frequently minor, as is expected
in a $\Lambda$CDM Universe, will stir galaxies up \citep[e.g.,][]{covi}.  (2.) The amount of mass accreted onto 
galaxies may be decreasing with time.  Mass accretion, although smoother
than mergers, might still disturb pre-existing disks \citep[see e.g.,][for a simple analytic treatment]{bour11,cacc}. 
(3.) Galaxies likely have larger molecular gas reservoirs at higher redshift \citep{daddi, tacc}. 
Higher gas fractions should result in more star-formation, and feedback from star-formation may also stir up the 
gas in galaxies \citep[e.g.,][and references therein]{silk}.  (4.) Finally, higher gas fractions can also lead to 
violent disk instabilities \citep[e.g.,][and references therein]{bour11,cacc} 
which increase random motions, and may also be a step in the direction of more star-formation.

At bottom, the primary causative factors of disk settling are likely two-fold:  more merging/accretion and higher
gas fractions at early times.  Faster star-formation rates are likely byproducts of these two factors, and increasing potential well
depths a byproduct of merging/accretion.
Since both factors decline with time, a general kinematic settling with time is expected.  Furthermore, both
factors are apparently declining {\it earlier} in massive galaxies,
at least since $z\sim1$, providing a natural explanation of the kinematic downsizing we find.

It is yet unknown whether disk galaxies in cosmological numerical simulations undergo a kinematic settling akin to what we
find in this paper.  Only recently have hydrodynamic simulations of galaxy formation been run with enough numerical resolution to study
the effects of mass accretion in detail.  Simulations find that mass accretes directly onto galaxies from cosmic 
filaments \citep[e.g.,][]{katz, birn,kere,ocvi,broo}.  The effects of accretion have been studied with 
small samples of simulated galaxies at $z\sim2$ \citep[e.g.,][]{bour09,ceve}.  These studies find that mass accretion perturbs disks and leads
to violent disk instabilities, but the effects of accretion have yet to be studied for a representative sample of simulated galaxies at lower redshifts.
The simulations which have been run to lower redshifts generally focus on reproducing the photometric and structural properties of galaxies
\citep[e.g.,][and references therein]{gove07, gove09, brook,mart}, and not $\sigma_g$, $V_{rot}$, and $S_{0.5}$.  On the analytic side, \citet{cacc} created a simple
model for the kinematic evolution of massive disk galaxies at $z\sim2$.  They predict that these galaxies, if still around 
today, are massive disks with similar rotation velocities but less disordered motions, similar to the kinematic settling we find.
Numerical simulations are not yet ripe for studies of the kinematic evolution of blue galaxies since $z\sim1$, and 
we look forward to comparisons with future predictions.

Our data are augmented with a compilation of measurements of the kinematics of emission-line galaxies from the literature over $0 < z < 3.8$.
Observations of local massive disk galaxies reveal few examples where disordered motions (as quantified by $\sigma_g$) play 
as an important role as in similar-mass emission-line galaxies at higher redshifts \citep[e.g.,][]{ghasp}.
Measurements of galaxy kinematics at $z>1.5$ show few clear
trends with redshift, likely primarily due to sample selection since only the most highly star-forming systems
at these redshifts can currently be studied, but also due to instrumental artifacts and observational
conditions.  However, it is clear that higher redshift galaxies have
a significant amount of disordered motions, whether or not they show evidence of rotation.

In summary, galaxies seem to have a life cycle:  Early in their lives they are accreting baryons rapidly, undergoing mergers, and possess
a large amount of gas.  Later in their lives the accretion decreases along with their gas content.  Kinematic
settling appears to be a natural, additional facet of this basic evolutionary arc.

\acknowledgments
S.A.K is supported by an appointment to the NASA Postdoctoral Program at
NASA's Goddard Space Flight Center, administered by Oak
Ridge Associated Universities through a contract with NASA.
The authors also acknowledge NSF grants AST 95-29098 and 00-71198 to UC Santa Cruz.
S. A. K. kindly thanks F. Bournaud, N. F\"{o}rster-Schreiber,
A. Gnerucci, M. Puech, P. van der Werf, and S. Wright for providing tables of measurements
and/or further information on how their measurements were performed.
We wish to extend thanks to those of Hawaiian ancestry on whose sacred mountain we are privileged guests.
MCC acknowledges support provided by NASA through Hubble Fellowship 
grant \#HF-51269.01-A, awarded by the Space Telescope Science Institute, 
which is operated by the Association of Universities for Research in 
Astronomy, Inc., for NASA, under contract NAS 5-26555. MCC also acknowledges 
support from the Southern California Center for Galaxy Evolution, a multi-campus
research program funded by the University of California Office of Research.

\end{document}